%% file: MAIN.tex
\documentclass[useAMS, referee, usenatbib]{biom}

\usepackage{natbib}
\usepackage{amsmath,amsfonts}
\usepackage{graphicx}
\usepackage{subfigure}
\usepackage{enumerate}
\usepackage{natbib}
\usepackage{url} 
\usepackage{booktabs}
\usepackage{tabularx}
\usepackage{lscape} 
\usepackage{bmpsize}
\usepackage{setspace}
\usepackage{mathrsfs}

\usepackage[hidelinks]{hyperref}
\hypersetup{
    colorlinks,
    linkcolor={red!50!black},
    citecolor={blue!50!black},
    urlcolor={blue!80!black}
}

\usepackage[utf8]{inputenc}
\usepackage{multirow,bigdelim, rotating}

\usepackage{zwcommands}
\usepackage{setspace}
\usepackage{mathtools}

\usepackage{dirtytalk}
\MakeRobust{\say}

\usepackage{xr}

\usepackage{hhline}
\usepackage{natbib}

\title[PAIR-GEE: Regression Analysis of Alternating Recurrent Events]{A Random Forest Inverse Probability Weighted Pseudo-Observation Framework for Alternating Recurrent Events}

\author{Abigail Loe$^{1,*}$\email{agloe@umich.edu}, 
Susan Murray$^{1}$, and Zhenke Wu$^{1}$ \\
$^1$: Department of Biostatistics, University of Michigan, Ann Arbor, Michigan, U.S.A.}

\begin{document}





\pagerange{\pageref{firstpage}--\pageref{lastpage}} 
\volume{16}
\pubyear{2025}
\artmonth{October}




\label{firstpage}


\begin{abstract}
Alternating recurrent events, where subjects experience two potentially correlated event types over time, are common in healthcare, social, and behavioral studies. Often there is a primary event of interest that, when triggered, initiates a period of treatment and recovery measured via a secondary time-to-event. For example, cancer patients can experience repeated blood clotting emergencies that require hospitalization followed by discharge, people with alcohol use disorder can have periods of addiction and sobriety, or care partners can experience periods of depression and recovery. Potential censoring of the data requires special handling. Overlaying this are the missing at-risk periods for the primary event type when individuals have initiated the primary event but not reached the subsequent secondary event.
In this paper, we  develop a framework for regression analysis of censored alternating recurrent events that uses a random forest inverse probability weighting strategy to avoid bias in the analysis of the time to the primary event due to informative missingness from the alternate secondary state. The proposed regression model estimates $\tau$-restricted mean time to the primary event of interest while taking into account complexities of censored.
Simulations show good performance of our method when the alternate times-to-event are either independent or correlated. We analyze a mobile health study data to evaluate the impact of self-care push notifications on the mental state of caregivers of traumatic brain injury patients.
\end{abstract}

%



\maketitle


%

\section{Introduction}
\label{s:intro}
\input{1Intoduction}
\section{Censored Longitudinal Data Structure}
\label{s:structure}
\input{2CensoredLongitudinalDataStructure}

\section{Construction of Pseudo-Observations}
\label{s:POs}
\input{5Pseudo}

\section{Inverse Probability Weighted Pseudo-Observations}
\label{s:mod}
\input{3Model}

\section{Estimation of Weights via Random Forest}
\label{s:RF}
\input{4RF}

\section{Simulation Study}
\label{s:SimStudy}
\input{6simStudy}
\section{Application to the CareQOL Study}
\label{s:data-app}
\input{7DataApplication}

\section{Discussion}
\label{s:discussion}
\input{8Discussion}

\backmatter


\section*{Acknowledgments}

We thank the investigators, coordinators, and research associates or assistants who worked on the CareQOL study (PI: Noelle E Carlozzi; \href{https://clinicaltrials.gov/study/NCT04570930}{NCT04570930}), the study participants, and the organizations who supported recruitment efforts. We also thank Jitao Wang for various assistance in data preprocessing and guidance. ZW is partly supported by a grant R01NR013658 from National Institute of Health, National Institute of Nursing Research.

\section*{Data Availability Statement}
The CareQOL data that support the findings in this paper are available upon reasonable request. The R code for reproducing the simulation and data analysis is accessible via \url{http://github.com/AbigailLoe/pair_gee}. 







\bibliographystyle{biom} \bibliography{biomsample}

\appendix




\label{lastpage}

\end{document}


\def\spacingset#1{\renewcommand{\baselinestretch}%
		{#1}\small\normalsize} \spacingset{1}
	
\if0\blind
	{		
\title{\bf Supplementary Materials for \\``A Random Forest Inverse Probability Weighted Pseudo-Observation Framework for Alternating Recurrent Events"}
		\author[1,$^\ast$]{Abigail Loe}
		\author[1,$^+$]{Susan Murray}
		\author[1,$^+$]{Zhenke Wu}
		\affil[1]{Department of Biostatistics, University of Michigan, Ann Arbor, MI 48109, USA; E-mail: $^\ast${\tt agloe@umich.edu}; $^+$: co-senior authors}
		
		\date{}
		\maketitle
} \fi
	
\if1\blind
	{
		\title{\bf Probability Weighted Pseudo-Observations for Alternating Recurrent Events}
		\author[]{}
		\date{}
		\maketitle
	} \fi
	
\vspace{-1cm}

\bigskip




\renewcommand{\thefigure}{S\arabic{figure}}
\renewcommand{\theequation}{S\arabic{equation}}
\renewcommand{\thetable}{S\arabic{table}}
\renewcommand{\thetheorem}{S\arabic{theorem}}
\renewcommand{\thesection}{A\arabic{section}}

\setcounter{figure}{0}
\setcounter{equation}{0}
\setcounter{table}{0}
\setcounter{theorem}{0}

\section{Clinical Relevance of $T_i(t)$}\label{s:supp:ClinRel}
\begin{figure}
\centerline{%
\includegraphics[width=.9\textwidth]{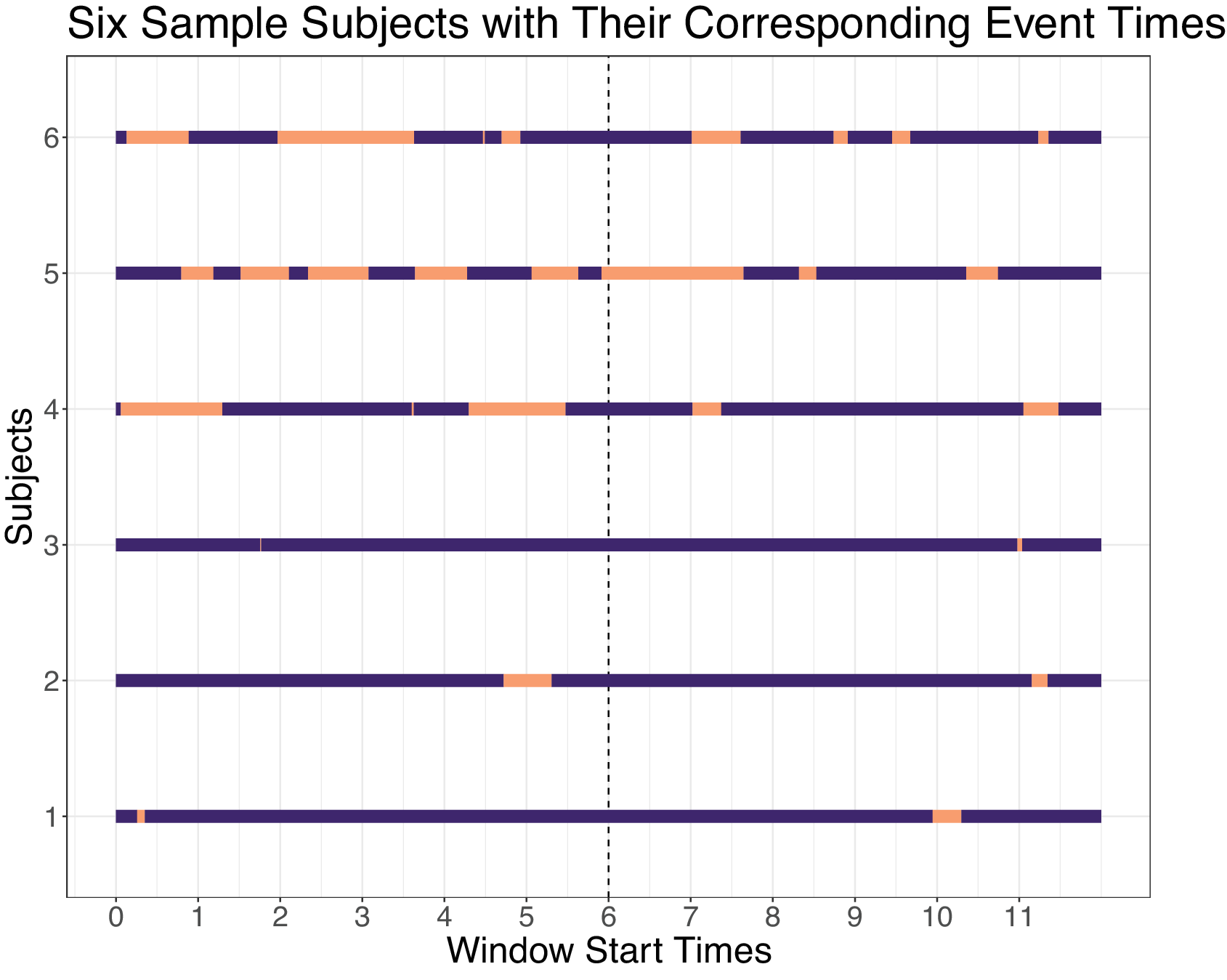}}
\caption{Six representative simulated subjects' alternating recurrent event data. Individuals $i = 1, 2, 3$ have $Z_i = 0$, and all other individuals have covariate profile $Z_i = 1$. Window start times, $t\in \mathcal{T}$, are displayed on the horizontal axis. Times at which subjects are at risk for the primary event are displayed in dark purple, and periods of the secondary event type, are displayed in light salmon. The time at which subjects experience a bend in hazards is distinguished with a vertical dashed line. Subjects $1$ through $3$ experience more time at risk relative to subjects $4, 5,$ and $6$. The latter subjects are more likely to be missing from the complete case analysis at window start times, $t \in \mathcal{T}$. }\label{supp:fig:sampSubj}
\end{figure}

Understanding how an exposure may increase or decrease time-to-primary event is a common question of clinical interest. Consider the scenario where $Z_i(t)$, dynamic predictors of alternating recurrent event behaviors contains an exposure of interest, and examine Figure \ref{supp:fig:sampSubj}. Subjects 4, 5, and 6 are at-risk rarely, and thus, appear in the censored longitudinal data structure infrequently. This by itself is not a problem, as our censored longitudinal data structure is constructed to handle such missingness, though if an exposure of interest shortens time spent at-risk, these individuals will be missing from the data, and their shorter times-to-primary alternating recurrent event will not contribute information to the proposed regression model. This will produce sampling bias, and may obscure potentially significant associations between exposures and outcomes. 

An analyst may also attempt a ``naive" approach to modeling the censored longitudinal data structure, and model $T_i ^{(1)}(t)$ rather than $T_i(t)$. One may notice that $T_i ^{(1)}(t)$ is always defined, whereas $T_i(t)$ is subject to missingness. That is, for subjects with $R_i(t)=1$, $T_i ^{(1)}(t)=T_i(t)$ but for subjects with $R_i(t) = 0$, $T_i^{(1)}(t) = \{T^{(1)}_{i,\eta_i^{(1)}(t)}- T^{(2)}_{i,\eta_i^{(2)}(t)}\} + T_i^{(2)}(t)$. Thus, changes in $T_i^{(1)}(t)$ for $R_i(t) = 0$ are a function of both the primary and secondary types of events, which cannot be untangled in a GEE approach. To understand why this is a problem, consider the example of cancer patients with venous thromboembolisms, where the primary event is time-to-hospitalization and the secondary alternating recurrent event is time-to-discharge from hospital. Imagine a certain kind of chemotherapy, Chemo A, is associated with the same time-to-hospitalization as Chemo B, but an increase in duration of hospital stay. For subjects with $R_i(t) = 1$ and Chemo A, $T_i^{(1)}(t)$ is the same as $T_j^{(1)}(t)$ for those with Chemo B. However, for subjects with $R_i(t) = 0$ and Chemo A, $T_i^{(1)}(t)$ is actually greater than $T_j^{(1)}(t)$; this may lead to an analyst concluding that Chemo A, the therapy that increases hospitalization time, is a superior treatment, as it increases average $T_i^{(1)}(t)$. However, the observed increase was due to an increase in the secondary alternating recurrent event, and in this situation is something that most clinicians would prefer to decrease.


\section{Proof of Theorem 1}\label{proofProp1}
\begin{align*}
&\EE\left\{\frac{R_i(t)}{\pi_i(t)} PO_i^\tau(t)\bigg| Z_i(t)\right\} =\EE\left\{\frac{R_i(t)}{P[R_i(t)=1|W_i (t)]} PO_i^\tau(t)|Z_i(t)\right\}\\ 
\intertext{But by conditioning on a finer partition, we obtain:}
&=\EE\left\{ \EE\left\{\frac{R_i(t)}{P[R_i(t)=1|W_i (t)]} PO_i^\tau(t)|Z_i(t),W_i(t), R_i(t)\right\} |Z_i(t)\right\}\\ 
\intertext{However, if we have conditioned on $W_i(t)$ and $R_i(t)$, then they may be treated as fixed in the inner expectation and factored, and we also apply Assumption \ref{TimodCorrect}:}
&=\EE\left\{\frac{R_i(t)}{P[R_i(t)=1|W_i (t)]} \EE\left\{ PO_i^\tau(t)|Z_i(t),R_i(t)\right\} |Z_i(t)\right\} 
\\
\intertext{Now, apply the definition of our pseudo-observation and expectation of a Bernoulli random variable to obtain}
&=\EE\left\{\frac{R_i(t)}{P[R_i(t)=1|W_i (t)]} R_i(t) \EE[\min \{T_i^{(1)}(t), \tau\}|R_i(t) = 1, Z_i(t)] |Z_i(t)\right\} 
\\
&=\EE\left\{\frac{R_i(t)}{P[R_i(t)=1|W_i (t)]}  \EE[\min \{T_i^{(1)}(t), \tau\}|R_i(t) = 1, Z_i(t)] |Z_i(t)\right\} 
\\
\intertext{However, if $R_i(t)=1$, then by definition $\min\{T_i^{(1)}(t), \tau\} = \min\{T_i(t), \tau\}$, and by Assumption \ref{TimodCorrect}, we can condition on $W_i(t)$:}
&=\EE\left\{\frac{R_i(t)}{P[R_i(t)=1|W_i (t)]}  \EE[\min \{T_i(t), \tau\}|R_i(t) = 1, Z_i(t),W_i(t)] |Z_i(t)\right\} 
\\
&=\EE\left\{  \EE\left[\frac{R_i(t)}{P[R_i(t)=1|W_i (t)]}\min \{T_i(t), \tau\}|R_i(t) = 1, Z_i(t),W_i(t)\right] |Z_i(t)\right\}
\\
\intertext{But marginalizing over $W_i(t)$ yields: }
&=\EE\left\{  \frac{R_i(t)}{P[R_i(t)=1|W_i (t)]}\min \{T_i(t), \tau\} |Z_i(t)\right\} 
\\
\intertext{The reasoning now boils down to a more classic argument involving inverse probability weights.}
&=\EE\left\{\frac{R_i(t)}{P[R_i(t)=1|W_i (t)]}  \min \{T_i(t), \tau\}|Z_i(t)\right\}\\ 
&=\EE\left\{ \EE\left\{\frac{R_i(t)}{P[R_i(t)=1|W_i (t)]}  \min \{T_i(t), \tau\} |W_i (t), Z_i(t) \right\}|Z_i(t)\right\} \\
\intertext{However, by conditioning on $W_i(t)$, we may again treat $P[R_i(t)=1|W_i (t)]$ as a constant:}
&=\EE\left\{ \frac{1}{P[R_i(t)=1|W_i (t)]}\EE\left\{R_i(t)  \min \{T_i(t), \tau\} |W_i (t), Z_i(t) \right\}|Z_i(t)\right\} \\
\intertext{By Assumption \ref{conditionalIndependence}:}
&=\EE\left\{ \frac{1}{P[R_i(t)=1|W_i (t)]}\EE\left\{R_i(t) |W_i (t), Z_i(t) \right\} \EE \left\{ \min \{T_i(t), \tau\} |W_i (t), Z_i(t) \right\}|Z_i(t)\right\} \\
&=\EE\left\{ \frac{1}{P[R_i(t)=1|W_i (t)]}P\left\{R_i(t) = 1 |W_i (t), Z_i(t) \right\} \EE \left\{ \min \{T_i(t), \tau\} |W_i (t), Z_i(t) \right\}|Z_i(t)\right\} 
\end{align*}
As $W_i(t)$ is assumed to contain covariates $Z_i(t)$ (Assumption \ref{conditionalIndependence}), the above line reduces to
$\EE\left\{ \min \{T_i(t), \tau\} |Z_i(t)\right\}$, as desired.

\section{History Covariates}\label{supp:histCovs}
When recurrent event times with negligible secondary alternating recurrent event length are highly correlated within an individual, \citet{Chapter1} found that ``history covariates," whether or not they are available for analysis at the beginning of follow-up, are some of the most important types of covariates included in $Z_i(t)$ for dynamic prediction of recurrent events. In this section we first develop history covariates for: (1) \textit{full} history setting (those where $Z_i(t)$ is known for all $t\in \mathcal{T}$) and (2) \textit{partial} history covariates (those where $Z_i(t)$ is not completely known at the start of follow-up), and develop a new type of history covariate for the prediction of $\hat{R}_i(t)$. For all defined covariates, we first consider the full history covariate setting, where past subject behavior within $t_h$ units of time prior to $t_0$ may be used in defining covariates, and then describe different cases of partial history information that a practitioner may encounter.



For subject $i$, let $q_i$ count the number of primary events that occur in the time period from $t_h$ to $t_0$. As subjects are assumed to be at-risk at $t_0= 0$, the events that occur prior to $t_0$ may be denoted $\max(T_{i, -(q_i+1)}^{(2)}, t_h) < T_{i, -q_i}^{(1)}< T_{i, -q_i}^{(2)}< \ldots T_{i, -1}^{(1)}< T_{i, -1}^{(2)}$, where calendar times for events happening prior to $t_0$ and $t_h$ are less than zero. First, consider a dynamic predictor that estimates the mean observed time at-risk for a primary alternating recurrent event for subject $i$. This covariate may be written as 
\begin{equation}\label{eq:avg_time_at_risk}
    H_{f, 1}^{(1)}(t) = \frac{1}{\sum_{j=-q}I(T_{i, j}^{(1)}<t)}\sum_{j=-q}\{T_{i,j}^{(1)}-\max (T_{i, (j-1)}^{(2)}, t_h)\}\times I(T_{i, j}^{(1)}<t).
\end{equation}
 This quantity is well defined for all $t \in \mathcal{T}$, and for later $t \in \mathcal{T}$, the quality of the information carried by $H_{f,1}^{(k)}$ increases due to an increased number of gap times used in averaging.  
 
 We also consider the most recent primary event time defined as \begin{equation}\label{eq:time_since_most_recent_hosp}
     H_{f, 2}^{(1)}(t) = t - \max(\{T_{i,j}^{(1)}| T_{i,j}^{(1)}< t, -q_i <j\}, t_h)
 \end{equation} and the length of last period at-risk for the primary event:
 \begin{equation}\label{eq:length_last_at_risk}
     H_{f, 3}^{(1)}(t) = T_{i, \eta_i^{(1)}(t)-1}^{(1)}- T_{i, \eta_i^{(2)}(t)-1}^{(2)}
 \end{equation} where $t_h$ is used if either of these quantities does not exist. Finally, $q_i+\eta_i(t) -1$ may be used as a historical covariate, counting the number of primary events occurring before time point $t$. For all of the history covariates described, we focus on using the primary alternating recurrent event time, but note that the secondary alternating recurrent event times may also be used to derive other dynamic predictors. 

While complete historical information on subjects yields the best results in simulation, we recognize that such information may not always be available. The first partial history setting generates the same possible covariates as the fully observed history setting, but restricts $q_i$ to be at most four. This means that for an alternating recurrent event history on a subject, within the time span from $t_h$ to $t_0$, at most four events will be used in constructing covariates that follow equations \ref{eq:avg_time_at_risk}- \ref{eq:length_last_at_risk}. We refer to covariates of this type as $\bH_{p_1}$; similarly covariates of type $\bH_{p_2}$ have $q_i$ equal to at most two. A third partial history setting may arise when a proportion of subjects have complete subject history $t_h$, and the remaining subjects have no recorded event history. In simulation, we refer to covariates arising from this historical setting as $\bH_{p_3}$. Moreover, an analyst may not have any historical event information, which results in only history covariates constructed from events on study ($\bH_{p_4}$), or simply not include any history covariates, $\bH_{n}$.

In considering updating history covariates, the random forest described in Section \ref{s:RF} is only concerned with predicting $R_i(t)$ at each time point; thus covariates that may perfectly predict $\min(T_i(t), \tau)$ are permissible for use in our random forest algorithm. Furthermore, an individual's tendency to enter into the risk-set at other windows may be an important covariate in fitting a model to $R_i(t)$, particularly if they rarely appear in the risk-set for the primary event-type or almost always appear in the risk-set for the primary event-type; that is $R_i(t_l) = 0$ for all $t \neq t_l$ or $R_i(t_l) = 1$ for all $t \neq t_l$ respectively. Therefore, we introduce a new way to consider historical information, called \textit{reflective} history, or $\bH_{p*}$. 

Consider the prediction of $R_i(t_1)$, or the prediction of risk set membership at the start of the second follow-up window in the $\bH_{p_4}$ setting. There are likely few observations of either type of recurrent event outcome. However, there is a rich set of covariates useful for prediction of behavior in the window starting at $t_1$; those are the events occurring after $t_1$, particularly in the absence of censoring. Note that we are not using future primary alternating recurrent event times to predict current primary alternating recurrent event times. We are simply using future, already observed primary and secondary event types to better fit a model for $R_i(t)$. Such settings commonly arise when analyzing data from clinical trials; all outcomes have been recorded for the entire follow-up at data lock, so future primary and secondary events have been fully observed. Indeed, this is an extension of an argument made by \citet{TchetgenTchetgen}, where missingness patterns across all time points was considered in maximum likelihood estimation with longitudinal data subject to nonmonotone dropout. In the $\bH_{p*}$ setting, an analyst keeps track of the number of observed events both prior to and subsequent to all $t\in \mathcal{T}$, and then calculates covariates of the type of Equations \ref{eq:avg_time_at_risk}- \ref{eq:length_last_at_risk} using events prior to $t$ if there are more events before the window start time, or events later than $t$ if the opposite is true. For window start-times where there are relatively few observations available for small $t$ and no censoring, we expect improved model performance in the $\bH_{p*}$ setting compared to the $\bH_{p_4}$ setting.

\section{Independent Event Times}\label{supp:simStudy:ind}
We assume that gap times between alternating recurrent events are independent and identically distributed with an exponential distribution, though primary and secondary event types may have different rate parameters, $\lambda_i$ and $\gamma_i$ respectively. \citet{Xia:StatMed:2019:regression} showed that for independent exponentially distributed gap-times, the time-to-first-event in a follow-up window for the censored longitudinal data structure remains exponentially distributed with rate $\lambda$. 
We first simulate $Z_1 \sim \text{Beta}(5, 1)$, $Z_2, Z_3\sim MVN(\vec{0}, 
\Sigma)$, where $\Sigma = \begin{pmatrix}
    1 & 0.2\\
    0.2 & 1
\end{pmatrix}$. The corresponding $\lambda_i$ that will generate exponential gap times that follow our model are taken as the solution of: 
$$
\EE\left [\min\{T_i(t), \tau \}\mid Z_i(t) \right]=\int_{0^+}^{\tau^-} y \lambda_i e^{-\lambda_iy}dy + \tau e^{-\lambda_i \tau} = \beta_0 + \beta_1 Z_{i1} + \beta_2 Z_{i2} + \beta_3 I(Z_{i3}\geq 0) \times Z_{i1},
$$
where $\beta_0 = 0.5$, $\beta_1= 0.25$, $\beta_2 = -0.4$ and $\beta_3 = 0.3$.
Combinations of covariates that generate hazards outside of the range of $\frac{1}{6}$ to $\frac{5}{8}$ for the primary alternating recurrent event-type will be discarded. The secondary alternating recurrent event-type has a hazard described by
$\gamma_i = \exp\{X_{i2} + \frac{1}{2}X_{i 3} + 10 X_{i6}^2 - 5 (Z_{i 1}+Z_{i3}) \}$, where $X_2 \sim N(1, .25)$, $X_3 \sim \chi^2_3$, and $X_6\sim Bern(.5)$. There are three included noise covariates $(X_1, X_4, X_5)$, but the inclusion of $Z_1$ and $Z_3$ in the generation of $\gamma_i$ induces light correlation between $\lambda_i$ and $\gamma_i$ ($\rho$ of approximately $-0.10$). 
Results, displayed in Table \ref{supp:tab:censRes}, are comforting. As with most longitudinal and recurrent event models, if events are independent of each other, all models work well, are relatively unbiased, and demonstrate appropriate asymptotic behavior. The distribution of point estimates in the presence of censoring are displayed in Figure \ref{fig:beta_violin_censored_independent}.

\section{Supplemental Tables and Figures}
\begin{table}\centering 
  \caption{A table displaying results from simulation in the case where primary recurrent events have correlation of $\rho = 0.8$, and secondary recurrent events are independent of each other. The hazard of primary alternating recurrent event completely determines the hazard of the alternating secondary recurrent event, no subjects are censored, and results come from 500 replicates of $n=750$ each. } 
  \label{tab:CorUncensoredRes} \small 
\begin{tabular}{@{\extracolsep{5pt}} lcrccccc} 
\\[-1.8ex]\hline 
\hline \\[-1.8ex] 
Model & Covariate &  $\hat\beta-\beta$ & $|\hat\beta-\beta|/\beta$ & Coverage & $\hat{SE}$ & ESD & $\hat{SE}$/ESD\\
\hline 
\hline \\[-1.8ex]
PAIR-GEE, with $\bH_f$ & $\hat\beta_1$ & -0.001 & 0.001 & 0.950 & 0.013 & 0.013 & 0.985 \\ 
 & $\hat\beta_2$ & -0.003 & 0.004 & 0.919 & 0.015 & 0.016 & 0.931 \\ 
 & $\hat\beta_3$ &  0.006 & 0.007 & 0.949 & 0.011 & 0.011 & 1.037 \\ 
 & $\hat\beta_4$ & -0.001 & 0.001 & 0.913 & 0.013 & 0.014 & 0.906 \\
 \hline
PAIR-GEE with $\bH_{p_1}$ & $\hat\beta_1$ &  0.001 & 0.002 & 0.936 & 0.014 & 0.014 & 0.951 \\ 
 & $\hat\beta_2$ & -0.004 & 0.006 & 0.913 & 0.015 & 0.016 & 0.912 \\ 
 & $\hat\beta_3$ &  0.004 & 0.005 & 0.947 & 0.011 & 0.011 & 1.028 \\ 
 & $\hat\beta_4$ & -0.006 & 0.008 & 0.881 & 0.012 & 0.014 & 0.904 \\ 
 \hline
PAIR-GEE, with $\bH_{p_2}$ & $\hat\beta_1$ &  0.000 & 0.001 & 0.937 & 0.013 & 0.014 & 0.949 \\ 
 & $\hat\beta_2$ & -0.008 & 0.012 & 0.887 & 0.014 & 0.015 & 0.921 \\ 
 & $\hat\beta_3$ &  0.004 & 0.005 & 0.953 & 0.011 & 0.011 & 1.019 \\ 
 & $\hat\beta_4$ & -0.008 & 0.011 & 0.846 & 0.012 & 0.014 & 0.897 \\ 
 \hline
PAIR-GEE, with $\bH_{p*}$ & $\hat\beta_1$ & -0.002 & 0.002 & 0.969 & 0.013 & 0.012 & 1.085 \\ 
& $\hat\beta_2$ & -0.003 & 0.004 & 0.931 & 0.015 & 0.016 & 0.965 \\ 
& $\hat\beta_3$ & -0.002 & 0.002 & 0.947 & 0.010 & 0.010 & 1.051 \\ 
& $\hat\beta_4$ & -0.021 & 0.028 & 0.510 & 0.011 & 0.012 & 0.945 \\ 
\hline
PAIR-GEE, with $\bH_{p_3}$ & $\hat\beta_1$ & -0.013 & 0.018 & 0.793 & 0.012 & 0.012 & 1.009 \\ 
 & $\hat\beta_2$ & -0.025 & 0.039 & 0.507 & 0.013 & 0.013 & 0.977 \\ 
 & $\hat\beta_3$ & -0.001 & 0.001 & 0.953 & 0.010 & 0.010 & 1.030 \\ 
 & $\hat\beta_4$ & -0.013 & 0.017 & 0.763 & 0.012 & 0.013 & 0.906 \\ 
 \hline
PAIR-GEE, with $\bH_{p_4}$ & $\hat\beta_1$ & -0.016 & 0.023 & 0.697 & 0.012 & 0.011 & 1.018 \\ 
 & $\hat\beta_2$ & -0.033 & 0.052 & 0.277 & 0.012 & 0.013 & 0.984 \\ 
 & $\hat\beta_3$ & -0.001 & 0.001 & 0.963 & 0.010 & 0.010 & 1.030 \\ 
 & $\hat\beta_4$ & -0.020 & 0.027 & 0.539 & 0.011 & 0.012 & 0.925 \\ 
 \hline
PAIR-GEE, with $\bH_n$ & $\hat\beta_1$ & -0.024 & 0.034 & 0.424 & 0.011 & 0.011 & 0.991 \\ 
 & $\hat\beta_2$ & -0.036 & 0.056 & 0.204 & 0.012 & 0.013 & 0.961 \\ 
 & $\hat\beta_3$ & -0.021 & 0.025 & 0.317 & 0.009 & 0.009 & 1.000 \\ 
 & $\hat\beta_4$ & -0.042 & 0.055 & 0.017 & 0.010 & 0.011 & 0.914 \\ 
 \hline
Unweighted GEE & $\hat\beta_1$ & -0.026 & 0.036 & 0.357 & 0.011 & 0.011 & 1.019 \\ 
 & $\hat\beta_2$ & -0.043 & 0.068 & 0.059 & 0.012 & 0.012 & 1.009 \\ 
 & $\hat\beta_3$ & -0.022 & 0.026 & 0.280 & 0.008 & 0.008 & 1.025 \\ 
 & $\hat\beta_4$ & -0.047 & 0.061 & 0.001 & 0.009 & 0.009 & 0.968 \\ 
\hline \\[-1.8ex] 
\end{tabular} 
\end{table}

\begin{table}\centering 
  \caption{A table displaying results from simulation in the case where primary recurrent events and secondary recurrent events are independent of each other. The hazard of primary alternating recurrent event completely determines the hazard of the alternating secondary recurrent event, approximately 30\% of subjects are censored, and results come from 300 replicates of $n=750$ each. } 
  \label{supp:tab:censRes} \small 
\begin{tabular}{@{\extracolsep{5pt}} lcrccccc} 
\\[-1.8ex]\hline 
\hline \\[-1.8ex] 
Model & Covariate &  $\hat\beta-\beta$ & $|\hat\beta-\beta|/\beta$ & Coverage & $\hat{SE}$ & ESD & $\hat{SE}$/ESD\\
\hline 
\hline \\[-1.8ex]
PAIR-GEE, with $\bH_f$ & $\beta_0$ & -0.001 & 0.002 & 0.957 & 0.034 & 0.035 & 0.980 \\ 
 & $\beta_1$ &  0.003 & 0.012 & 0.950 & 0.052 & 0.052 & 0.987 \\ 
 & $\beta_2$ &  0.000 & 0.001 & 0.937 & 0.036 & 0.038 & 0.947 \\ 
 & $\beta_3$ & -0.002 & 0.008 & 0.950 & 0.037 & 0.038 & 0.974 \\ 
 \hline 
PAIR-GEE, with $\bH_{p_1}$ & $\beta_0$ & -0.001 & 0.002 & 0.953 & 0.034 & 0.034 & 0.983 \\ 
 & $\beta_1$ &  0.003 & 0.012 & 0.957 & 0.051 & 0.052 & 0.985 \\ 
 & $\beta_2$ &  0.000 & 0.001 & 0.933 & 0.036 & 0.038 & 0.947 \\ 
 & $\beta_3$ & -0.002 & 0.008 & 0.947 & 0.037 & 0.038 & 0.974 \\ 
 \hline 
PAIR-GEE, with $\bH_{p_1}$ & $\beta_0$ & -0.001 & 0.002 & 0.957 & 0.034 & 0.034 & 0.983 \\ 
 & $\beta_1$ &  0.003 & 0.012 & 0.950 & 0.052 & 0.052 & 0.987 \\ 
 & $\beta_2$ &  0.000 & 0.000 & 0.933 & 0.036 & 0.038 & 0.947 \\ 
 & $\beta_3$ & -0.002 & 0.008 & 0.947 & 0.037 & 0.038 & 0.974 \\ 
 \hline 
PAIR-GEE, with  $\bH_{p*}$ & $\beta_0$ &  0.004 & 0.008 & 0.947 & 0.034 & 0.035 & 0.984 \\ 
 & $\beta_1$ &  0.000 & 0.002 & 0.947 & 0.052 & 0.053 & 0.986 \\ 
 & $\beta_2$ &  0.006 & 0.014 & 0.923 & 0.036 & 0.038 & 0.953 \\ 
 & $\beta_3$ & -0.006 & 0.021 & 0.953 & 0.037 & 0.038 & 0.978 \\ 
 \hline 
PAIR-GEE, with $\bH_{p_3}$ & $\beta_0$ & -0.001 & 0.002 & 0.953 & 0.034 & 0.034 & 0.983 \\ 
& $\beta_1$ &  0.003 & 0.013 & 0.947 & 0.052 & 0.052 & 0.987 \\ 
& $\beta_2$ &  0.000 & 0.000 & 0.933 & 0.036 & 0.038 & 0.950 \\ 
& $\beta_3$ & -0.002 & 0.008 & 0.950 & 0.037 & 0.038 & 0.974 \\ 
\hline 
PAIR-GEE, with $\bH_{p_4}$ & $\beta_0$ & -0.001 & 0.002 & 0.953 & 0.034 & 0.035 & 0.980 \\ 
 & $\beta_1$ &  0.003 & 0.012 & 0.947 & 0.052 & 0.052 & 0.985 \\ 
 & $\beta_2$ &  0.000 & 0.000 & 0.930 & 0.036 & 0.038 & 0.950 \\ 
 & $\beta_3$ & -0.002 & 0.008 & 0.953 & 0.037 & 0.038 & 0.974 \\ 
 \hline 
PAIR-GEE, with $\bH_n$ & $\beta_0$ & -0.001 & 0.002 & 0.957 & 0.034 & 0.035 & 0.980 \\ 
 & $\beta_1$ &  0.003 & 0.012 & 0.947 & 0.051 & 0.052 & 0.983 \\ 
 & $\beta_2$ &  0.000 & 0.001 & 0.927 & 0.036 & 0.038 & 0.949 \\ 
 & $\beta_3$ & -0.002 & 0.008 & 0.947 & 0.037 & 0.038 & 0.976 \\ 
 \hline 
Unweighted GEE & $\beta_0$ & -0.001 & 0.002 & 0.947 & 0.034 & 0.035 & 0.977 \\ 
 & $\beta_1$ &  0.003 & 0.011 & 0.947 & 0.052 & 0.053 & 0.974 \\ 
 & $\beta_2$ &  0.000 & 0.000 & 0.930 & 0.036 & 0.037 & 0.960 \\ 
 & $\beta_3$ & -0.002 & 0.008 & 0.943 & 0.037 & 0.037 & 0.992 \\ 
\hline \\[-1.8ex] 
\end{tabular} 
\end{table}

\begin{figure}
\centerline{%
\includegraphics[width=.99\textwidth]{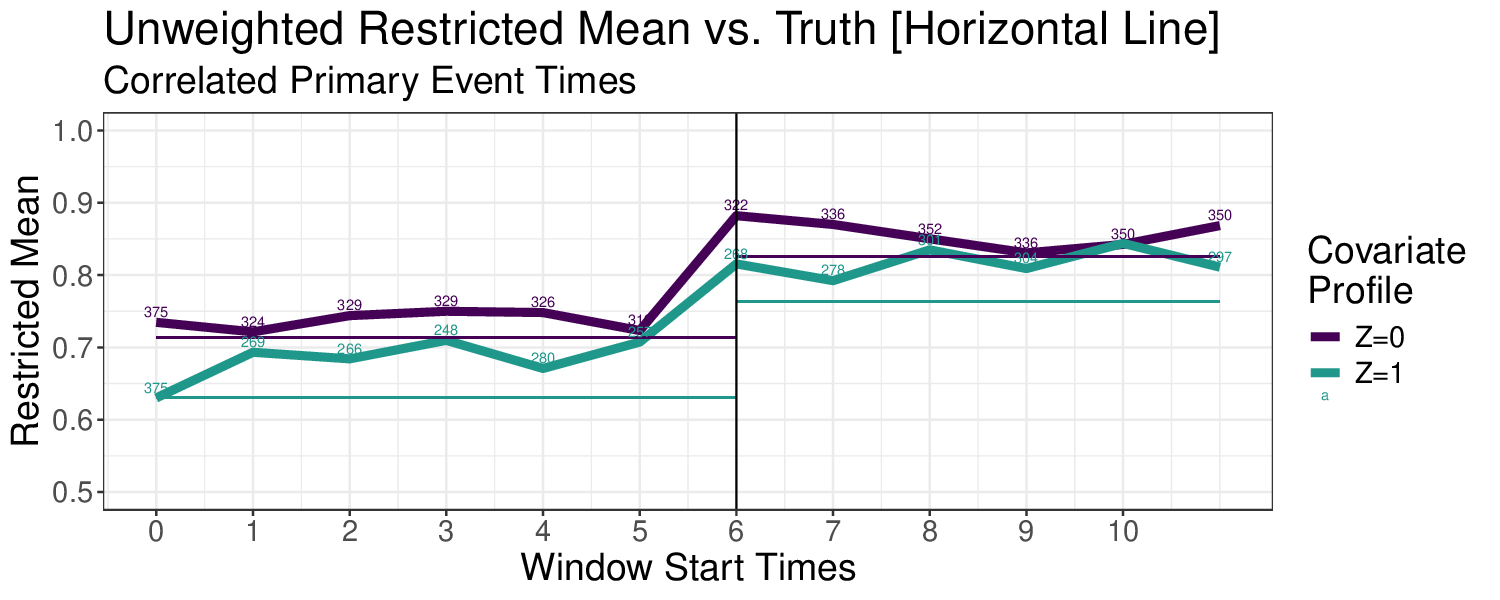}}
\caption{A graphic displaying the true $\tau$-restricted mean compared to the empirical, or observed, $\tau$-RMST in the highly correlated case for a single replicate. True parameter estimates are displayed using horizontal lines, while the empirical value is displayed on the $y$ axis. Lines connecting the estimated $\tau$-RMST are annotated with the number of subjects in the risk set for each window start time and each covariate profile.}\label{supp:fig:simReplicate}
\end{figure}

\begin{figure}
    \centering
    \includegraphics[width=\linewidth]{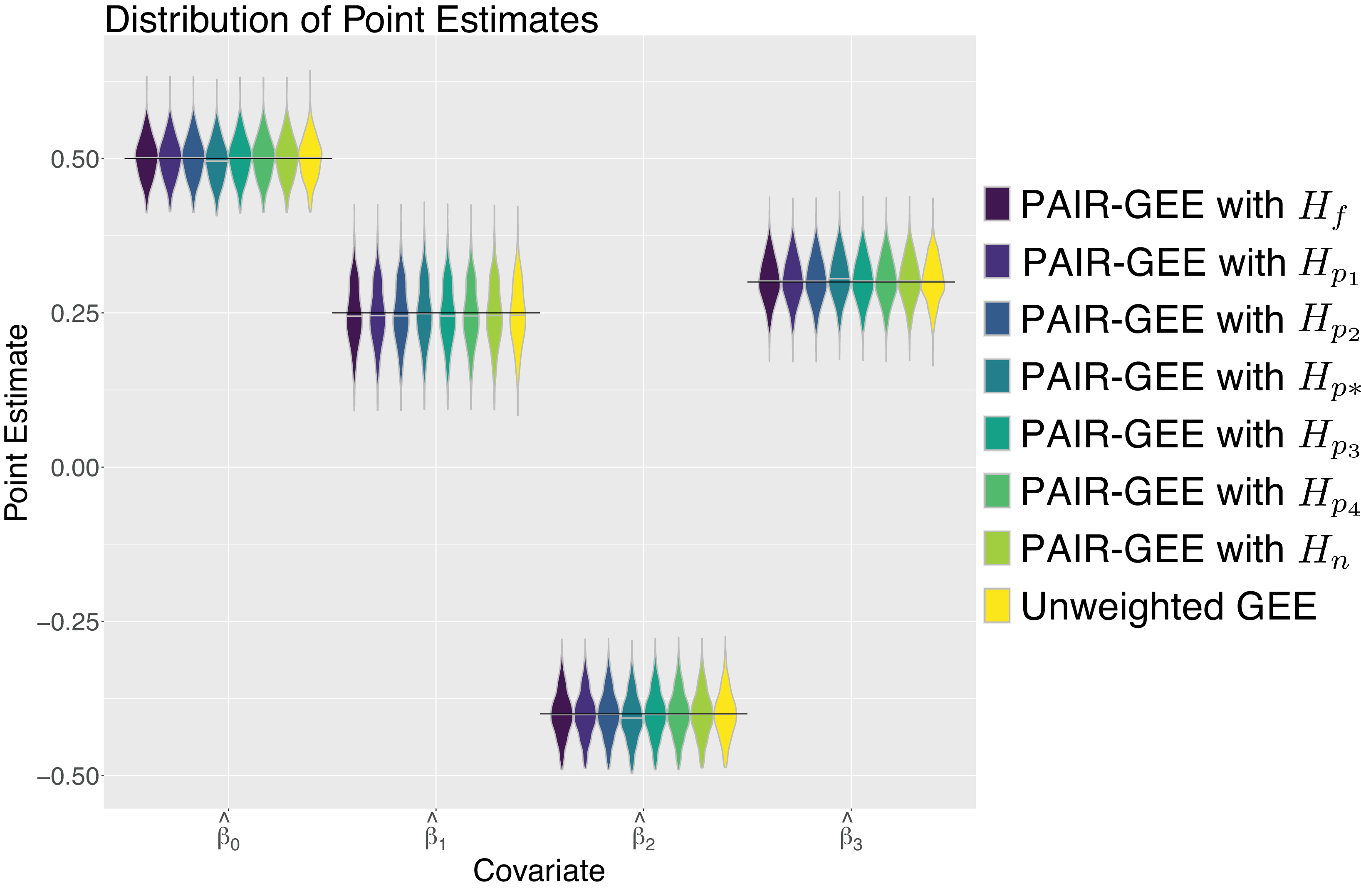}
    \caption{Distribution of point estimates in the presence of censoring, when primary alternating recurrent events within a subject are independent. Results come from 300 replicates of $n=750$ each. Features included in the random forest algorithm for estimating PAIR-GEE weights involve the seven history covariates described in Section \ref{s:RF} with varying degrees of availability that decreases from left to right for each displayed coefficient}
    \label{supp:fig:beta_violin_censored_independent}
\end{figure}

\begin{figure}
    \centering
    \includegraphics[width=0.9\linewidth]{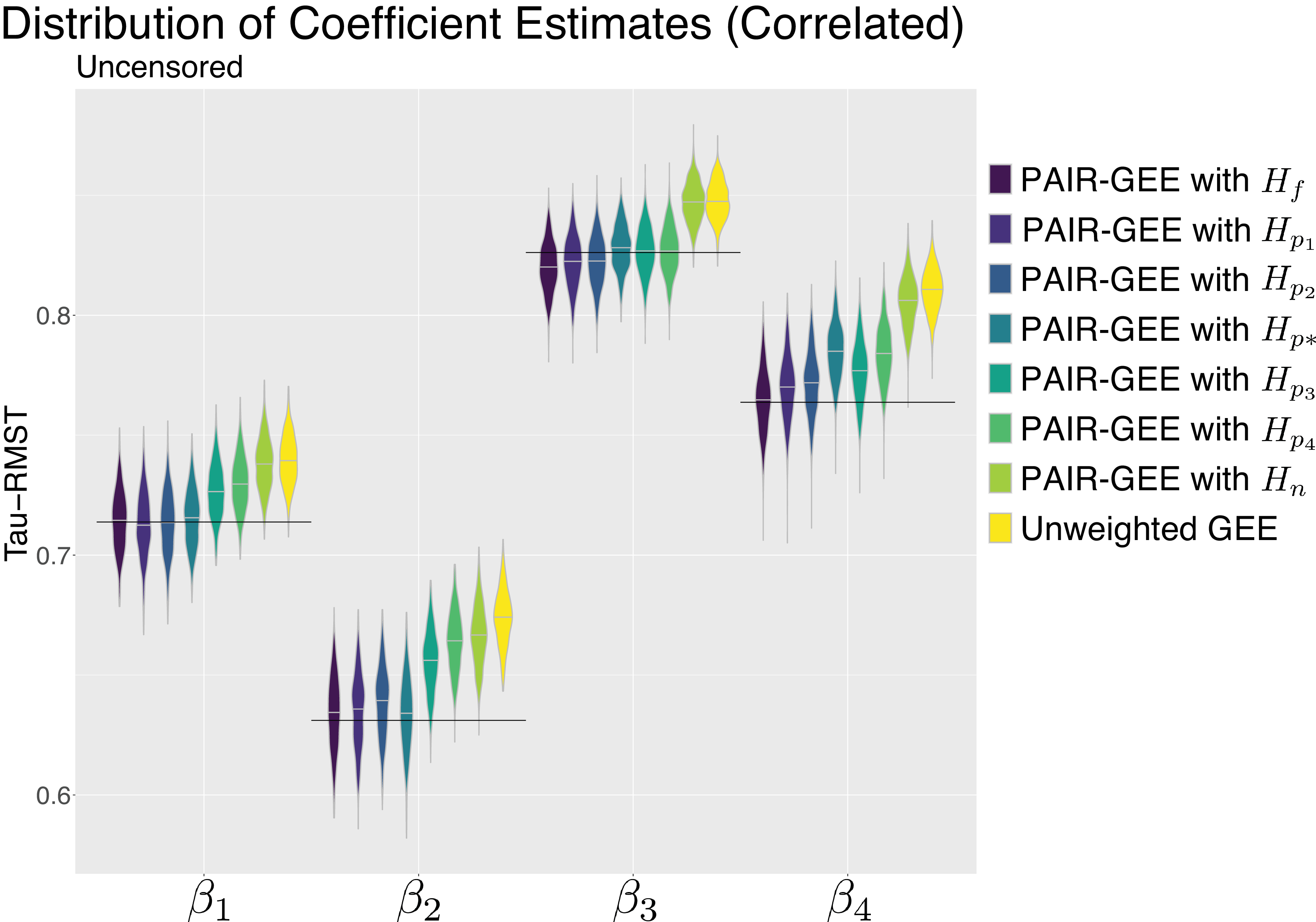}
    \caption{Violin plots of uncensored point estimates seen in simulation ($500$ replicates, $n=750$). Coefficients $\beta_1$ and $\beta_2$ reflect those with coefficients $Z_i=1$ and $Z_i = 0$ before $t = 6$ and coefficients $\beta_3$ and $\beta_4$ after and including $t = 6$, respectively. Features included in the random forest algorithm for estimating PAIR-GEE weights involve the seven history covariates described in Section \ref{s:RF} with varying degrees of availability that decreases from left to right for each displayed coefficient. Here, $\bH_f$ indicates the full year of patient history was available prior to $t=0$. $\bH_{p_1}$ and $\bH_{p_2}$ indicate that the patient history included information on the last two or one alternating recurrent events, respectively, if these occurred within the prior year. $\bH_{p*}$ indicates that no historical information prior to $t=0$ was available and that reflexive history covariates (as described in Section \ref{s:RF}) were used. $\bH_{p_3}$ indicates that complete history was available on 50\% of individuals, but was otherwise unavailable. $\bH_{p_4}$ indicates no availability of history covariates prior to $t=0$. $\bH_n$ indicates no history covariates were included in the random forest algorithm.}
    \label{supp:fig:uncensored_beta_plot}
\end{figure}

\begin{figure}
    \centering
    \includegraphics[width=\linewidth]{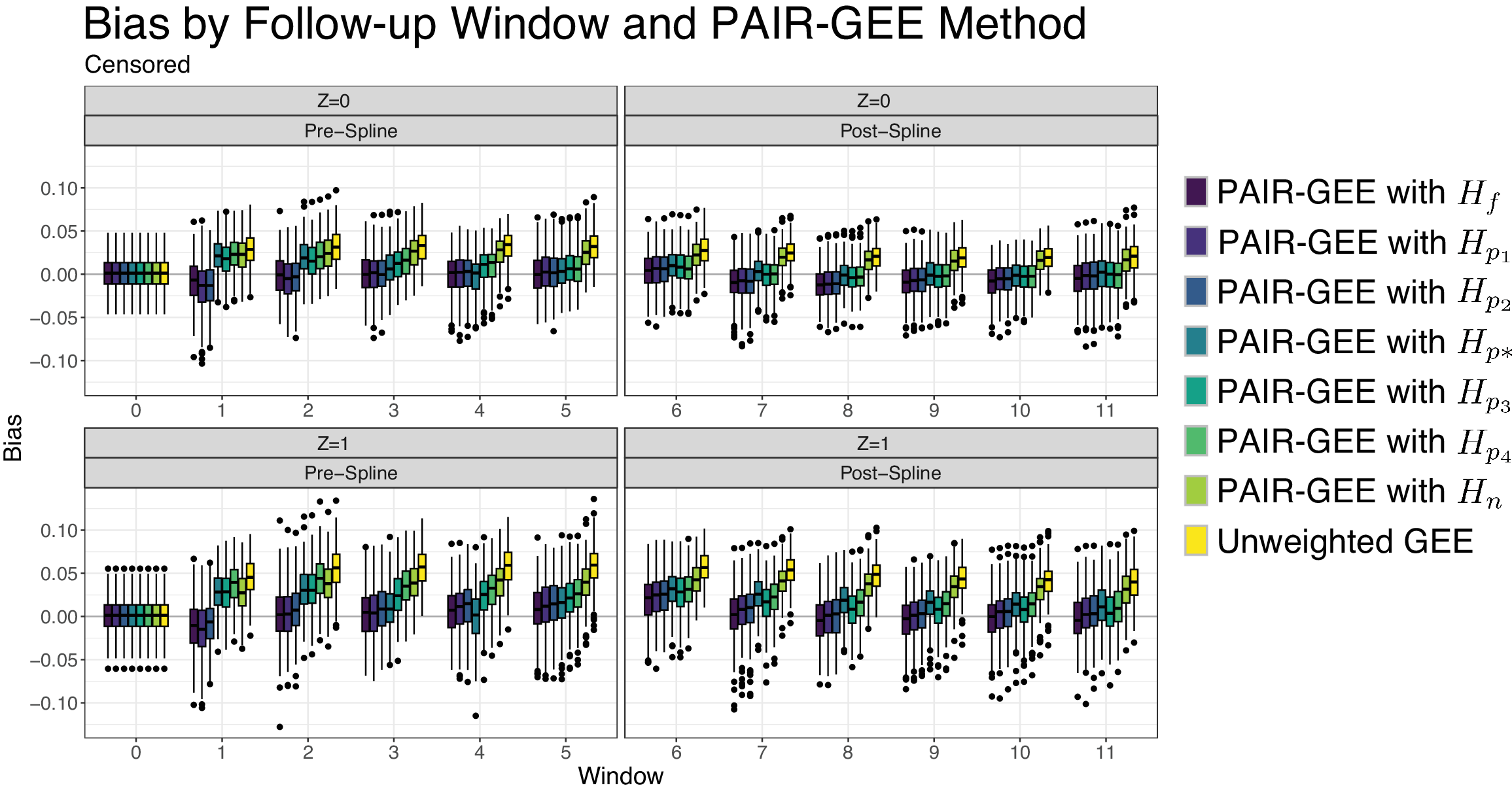}
    \caption{Bias in point estimation for each $t\in \mathcal{T}$ when alternating recurrent events are correlated with each other. The top half of the graphic displays results from the uncensored simulation, with the bottom half displaying results from the censored results, described in the main paper. Notice the persistent bias for the unweighted methods, as well as the PAIR-GEE with $\bH_n$, while other methods of incorporating history covariates are able, to varying extents, adjust for patient specific history in a more elegant manner, and approach non-zero bias in later follow-up windows.}
    \label{supp:fig:win_by_win_bias}
\end{figure}



%% file: 1Intoduction.tex
Recurrent events are common across multiple disciplines of research, such as economic, medical, behavioral, political, and biological. For example, in the biomedical context, patients with chronic illnesses often experience periods of relative health, punctuated by episodic instances of illness. Specifically, those with chronic obstructive pulmonary disease (COPD) experience breathing emergencies requiring hospital stays of varying duration.
Many statistical challenges for this data structure have been identified and addressed, including dependent censoring when using a gap-time analysis, and addressing correlation between recurrent event times \citep[e.g.,][]{  PropMeansModel,  frailtySmoking, frailtyRecid, XiaMurray:Biostat:2019:commentary, Xia:StatMed:2019:regression,KevinLiliDoug, zhao2020incorporating, biVec, Chapter1}.
In the COPD example, if times-to-release from hospitalization are long, they can be viewed as alternating times-to-event that occur after each hospitalization.
Alternating recurrent events of short duration do not have much impact on analyses of the primary recurrent event of interest (such as time-to-hospitalization), and have been largely ignored in the literature. However, longer alternating event durations can have profound impacts on the analysis of the time to the primary recurrent event of interest, which is the focus of our current manuscript. 

More formally, we define alternating recurrent events as a time-to-event of primary interest followed by a time-to-event indicating when the individual returns to the at-risk state for the primary event of interest. 
For example, in the CareQOL study \citep{CareQoLApp}, caregivers of patients with traumatic brain injuries (TBIs) experienced repeated periods of depression before returning to baseline. To truly understand the primary recurrent event of interest, i.e. time-to-depressed state, one needs to account for the time-to recovery from depression, i.e. the secondary (alternating) recurrent event type. A key concept that motivates our work is that subjects cannot be at-risk for entering a depressed state while currently in one. This missingness pattern related to the primary recurrent event time results in a biased analysis sample,  similar to that seen in \citet{RaoSamplingBias, RaoWildlife}, and \citet{OGLengthBias}. That is, since depression-prone patients are more likely to be removed from the risk set for the primary endpoint at regular intervals they are not adequately represented in the analysis.

This manuscript explores censored recurrent event analyses where alternating (secondary) event times introduce such bias. Our approach transforms the recurrent event data structure to a censored longitudinal dataset comparable to that of \citet{TayobMurray:Biostat:2015} and incorporates inverse probability weights (IPW) to address this sampling bias due to the secondary alternating recurrent events. The IPW weights are taken from a random forest model which allows for complex relationships between many potentially collinear predictors and removal from the risk set for the primary recurrent event of interest due to the secondary alternating recurrent event.
We also borrow ideas from pseudo-observation literature \citep{andersen2003generalised, andersen2004regression} to address the additional independent censoring mechanism in these settings. 

The rest of this paper is organized as follows. In Section \ref{s:structure}, we transform alternating recurrent event data into a censored longitudinal data format, based on which we construct pseudo-observations (PO) in Section \ref{s:POs}. Although the PO construction unlocks access to popular regression methods, it is subject to selection into the at-risk sets, which we correct using IPW in Section \ref{s:mod}. In Section \ref{s:RF}, we describe the random forest approach for obtaining probability weights and useful history covariates that substantially improve the performance of our method. Performance of the proposed method is evaluated in Section \ref{s:SimStudy}. The model is applied to mobile health data from the CareQOL study in Section \ref{s:data-app}, and we conclude with a discussion in Section \ref{s:discussion}.

%% file: 2CensoredLongitudinalDataStructure.tex
For individual $i$, alternating recurrent event times $T_{i, 0}^{(1)}< T_{i, 0}^{(2)}< T_{i, 1}^{(1)}<T_{i, 1}^{(2)}<\ldots$ are potentially censored by the random variable $C_i$, where $C_i$ is independent of $T_{i,j}^{(1)}, T_{i,j}^{(2)}$, $i = 1, ..., n$, and $j \geq 0$. Here, $j$ indexes the event pairings, $i$ indexes subjects, and the superscript $k$, $k\in \{1, 2\}$ indexes the alternating primary ($k =1$) events and secondary ($k=2$) events. 
Upon experiencing an event at $T_{i,j}^{(1)}$, individual $i$ is not in the set of individuals at-risk for the primary recurrent event time for $t\in [T_{i,j}^{(1)},T_{i,j}^{(2)})$ until the secondary alternating recurrent event occurs. That is, $T_{i,j}^{(2)}, j \geq 0$ are  times at which individual $i$ re-enters the population of individuals at-risk for the primary alternating recurrent event type.  
For instance, in the setting of TBI caregivers, $T_{i,j}^{(1)}, j \geq 0$ would be times when caregivers enter a depressive state, and $T_{i,j}^{(2)}, j \geq 0$ are the times when subjects leave the depressive state to be at-risk for re-entering a depressive state.

\begin{figure}
\centerline{%
\includegraphics[width=.9\textwidth]{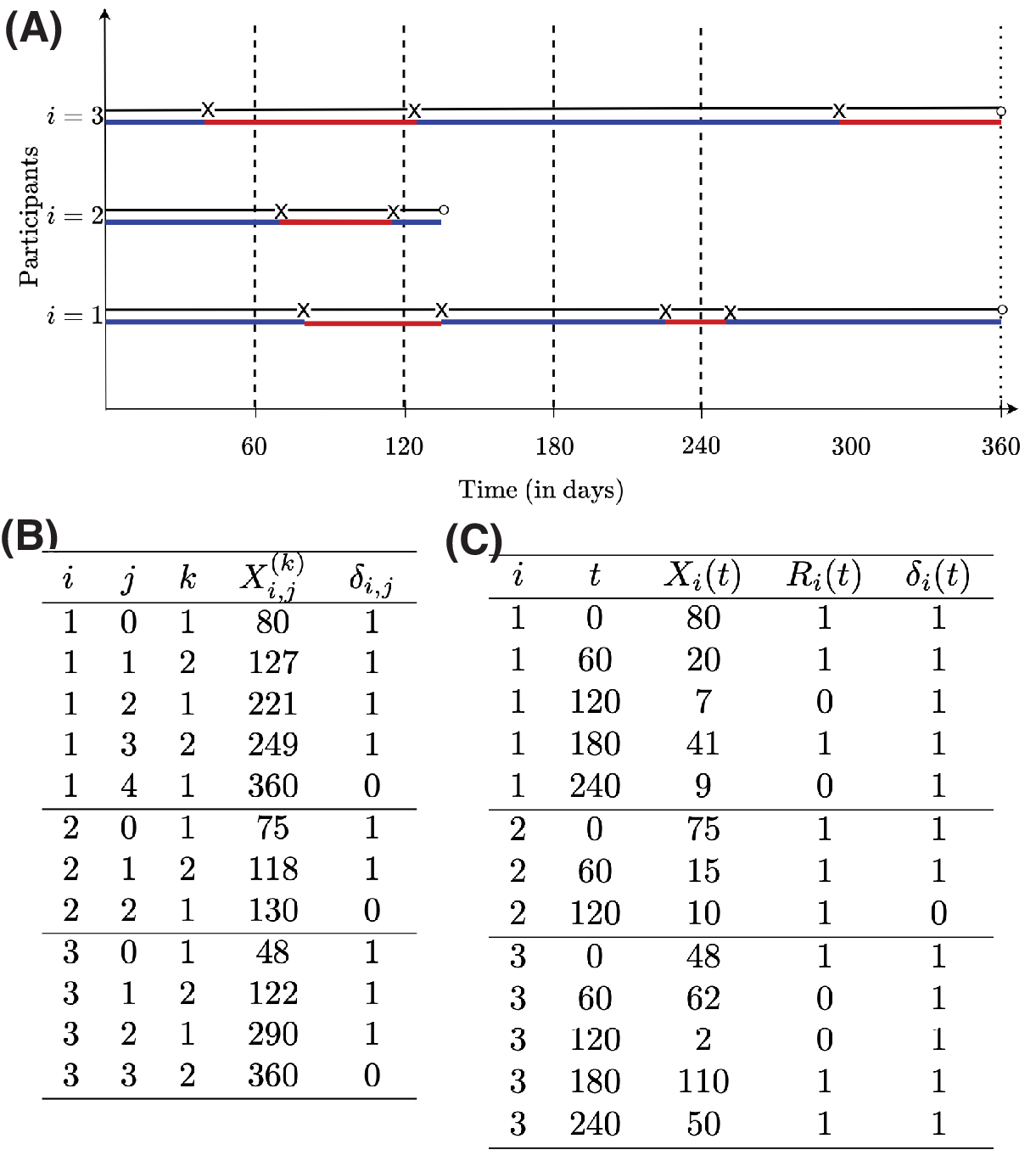}}
\caption{A visualization of alternating recurrent events in both the longitudinal sense and the alternating censored longitudinal format. In Panel (A), events within a subject are marked with an ``\texttt{X}." The primary and secondary event types are the start and the end of the {\color{red}red} segments, respectively. {\color{blue}Blue} segments represent at-risk periods for the primary event type. Start times of potential follow-up windows of length $\tau = 120$ corresponding to $\mathcal{T} = \{0, 60, 180, 240\text{ days}\}$ are represented with dashed vertical lines. Traditional observed alternating recurrent events are recorded in Panel (B), while Panel (C) corresponds to those same event converted to an alternating censored longitudinal format. All participants were administratively censored after $360$ days of follow-up, thus $t= 300$ is not included in $\mathcal{T}$. Note that $X^{(1)}_{1, 0}$ (the first `\texttt{X}' in subject 1) produces multiple measures in the alternating censored longitudinal data format. Additionally, subject 2 produces three measures where $R_2(t) =1$; subject 3 produces the same number of measures where $R_3(t)=1$, but was removed from the at-risk population at $t=60,120$.} \label{fig:data_viz}
\end{figure}

Due to potential censoring, observed data for each individual is a set of event times and indicator functions. In this manuscript, we denote potentially censored event times using $X_{i, j}^{(k)}= \min(T_{i,j}^{(k)}, C_i)$ and the corresponding censoring indicator $\delta_{i, j}^{(k)} = I(X_{i, j}^{(k)}< C_i)$. To avoid the well-known dependent censoring problem due to gap-time analysis, we follow the censored longitudinal data approach of \citet{TayobMurray:Biostat:2015}, \citet{Xia:StatMed:2019:regression}, \citet{zhao2020incorporating}, and \citet{Chapter1}. They transform correlated recurrent event data into a series of times-to-first-event in follow-up windows of length $\tau$, starting at $t\in \mathcal{T}_i = \{t_0, t_1, ... t_{b_i}\}$. When data are uncensored, $t_{b_i}$ is chosen so that $t_{b_i}+ \tau$ does not exceed available follow-up time. When observation $i$ is censored, $t_{b_i}= \max\{t_l \in \mathcal{T}_i: t_l <C_i\}$. For the follow-up window starting at time $t$, $n(t)$ is the number of uncensored subjects available for follow-up at $t$. See Figure \ref{fig:data_viz}, Panel (A) for a visualization of traditionally recorded alternating recurrent events with corresponding data in Panel (B). 

To create the censored longitudinal dataset, for each $t \in \mathcal{T}_i$, and each individual $i = 1,2, ...,n(t)$ who remains uncensored at $t$, it is helpful to map the relationship between the original alternating recurrent events denoted by $T_{i, j}^{(1)}< T_{i, j}^{(2)}$ for $j\geq 0$ with those that appear first in the follow-up window starting at $t$. The algorithm then becomes, for each individual $i = 1, 2, \ldots n(t)$ uncensored at time $t$:
\begin{itemize}
    \item[1)] Identify the $j$ index of the first alternating recurrent event of type $k = 1,2$ after $t$ by $\eta_i^{(k)}(t)= \min\{j|T_{i,j}^{(k)}> t, j = 0, 1, \ldots \}$. 
    \item[2)] Define residual censoring time from time $t$ (the start of the follow-up-window) as $C_i(t) = C_i - t$. Let $T_i^{(k)}(t) = T_{i, \eta^{(k)}_{i}(t)}^{(k)}- t$. Let  $X_i(t) = \min\{T_{i}^{(1)}(t),T_i^{(2)}(t), C_i(t) \}$ be the first observed event time measured from time $t$ regardless of its event type, with corresponding censoring indicator, $\delta_i(t) = I\{X_i(t) < C_i(t)\}$. 
    \item[3)] Define 
    $R_i(t) = I[T_{i}^{(1)}(t) < T_i^{(2)}(t)]$, to indicate that individual $i$ was at-risk for the primary alternating recurrent event as of time $t$.
    
    \item[4) ] Construct the censored longitudinal dataset at $t$ by $\big\{X_i(t), R_i(t), \delta_i(t), Z_i(t),  i = 1, 2, \ldots n(t) \big \}$ where $Z_i(t)$ are covariates that are known as of time $t$. The conversion of alternating recurrent events into a censored longitudinal framework is shown in Figure \ref{fig:data_viz}, Panel (C).
\end{itemize}


In this manuscript, we define $T_i(t)$ as the time to the primary recurrent event of interest if patients were at-risk for this event at all start-times, $t \in \mathcal{T}_i$. In the setting with alternating recurrent events, we only partially observe $T_i(t)= R_i(t) \times T_i^{(1)}(t)$ when  $R_i(t)=1$. When $R_i(t)= 0$, individual $i$ is not at-risk at time $t$ for this event, so that $T_i(t)$ is missing. 
For further discussion on the clinical relevance of $T_i(t)$ and consideration of $T_i(t)$ rather than $T_i^{(1)}(t)$, see Supplemental Materials Section \ref{s:supp:ClinRel}.
We will hereafter refer to $\mathcal{CC}(t)=\{ i = 1, 2, \ldots, n_R(t): R_i(t) = 1\}$ as the set of uncensored individuals at time $t$ with complete data on the primary recurrent event endpoint, with $n_R(t)\leq n(t)$ being the number of individuals in the set $\mathcal{CC}(t)$. We define 
$\{X_i(t), \delta_i(t): i \in \mathcal{CC}(t)\}$ as the complete case data for the primary recurrent event of interest at $t$. 
In Section \ref{s:mod}, we express our interest in understanding $T_i(t)$ for $i =1, 2, \ldots,  n(t)$, rather than $T^{(1)}_i(t)$ for $i \in \mathcal{
CC}(t)$, and define models with this goal in mind. In particular, we introduce longitudinal restricted mean regression models to estimate $E[\min \{T_i(t), \tau\}|Z_i(t)]$, where $\tau$ is the follow-up window length from $t$. The $\tau$-restricted mean time-to-event uses $\tau$ to focus inference on clinically relevant horizons, e.g., 30-day readmission in hospital outcomes or 5-year survival in oncology, matching clinical or policy relevance \citep[e.g.,][]{royston2013restricted}. But before details for model development are given, we summarize a pseudo-observation approach to address the censoring by $C_i(t)$ in Section \ref{s:POs}.


%% file: 5Pseudo.tex
\citet{andersen2003generalised} proposed a technical device now employed in many longitudinal survival contexts called \say{pseudo-observations,} a variation of jack-knife methods that provide asymptotically unbiased estimates of regression parameters in the presence of censoring. We provide a primer on the use of pseudo-observations for a particular follow-up window at time $t$, based on the $n_R(t)$ individuals in $\mathcal{CC}(t)$. 
The issue of selection into the risk set at $t$ caused by analyzing only individuals in $\mathcal{CC}(t)$ will be corrected in Section \ref{s:mod}.
We define $S_t^{(1)}(u)=P\{T_i^{(1)}(t)>u\}$ as the primary-event-free survival function based on the $n_R(t)$ individuals in $\mathcal{CC}(t)$, and $\theta(t)= E[\min\{T_i^{(1)}(t),  \tau\}|R_i(t) = 1] =\int_0^{\tau}S_t^{(1)}(u) du$, i.e., the complete case $\tau$-restricted mean event time at follow-up window $t$. 
A consistent nonparametric estimator for $\theta(t)$ is $\hat{\theta}(t) = \int_0^\tau \hat{S}_t^{(1)}(u) du,$ where $\hat{S}_t^{(1)}(u)$ is the Kaplan-Meier survival curve estimated from data pairs $\{X_i(t),\delta_i(t): i \in \mathcal{CC}(t)\}$.
Define leave-one-out estimator $\hat{\theta}^{-i}(t)$ as the estimator where individual $i$ is excluded from the cohort. 
Then a pseudo-observation for subject $i, i \in \mathcal{CC}(t)$ at time $t \in \mathcal{T}_i$ is:
$$
PO_i^\tau(t) = n_R(t) \hat{\theta}_i(t) - \{n_R(t) -1\} \hat{\theta}^{-i}(t),
$$
which can be easily obtained using the \texttt{pseudomean} function from the \texttt{survival} package in the programming language \texttt{R}. 
\citet{andersen2004regression} showed that pseudo-observations such as $PO_i^\tau(t)$, have the same conditional expectation, $E[PO_i^\tau(t)|R_i(t) = 1, Z_i(t)]$, as $E[\min \{T_i^{(1)}(t), \tau\}|R_i(t) = 1, Z_i(t)]$.
Essentially, once calculated, a pseudo-observation can be used in a regression analysis as an uncensored outcome as if one were actually modeling $E[\min\{T_i^{(1)}(t), \tau\}|R_i(t)=1, Z_i(t)]$. When calculating $PO^\tau_i(t)$ for $t\in \mathcal{T}$ and $i = 1, 2, \ldots, n_R(t)$, one obtains a longitudinal dataset that may be used with traditional longitudinal analysis methods.

%% file: 3Model.tex
In Section \ref{s:POs}, we constructed pseudo-observations, $PO_i^\tau(t)$, that have conditional expectation, $E[\min\{ T_i^{(1)}(t), \tau\}| R_i(t) = 1, Z_i(t)]$, so that an uncensored outcome is available for analysis, but it is still subject to selection into the risk set at time $t$, i.e., this does not necessarily correspond to $E[\min \{T_i(t), \tau\}\mid Z_i(t)]$.
Now using pseudo-observations, $PO_i^\tau(t)$, we show how inverse probability weights (IPW) may be used to correct this issue to construct models for estimating $E[\min\{ T_i(t), \tau\}|Z_i(t)]$. We need two assumptions for establishing the correction in Theorem 1.


\begin{assumption}[Ignorability]\label{conditionalIndependence}

    There exist covariates $W_i(t)$ such that $R_i(t) \perp\!\!\!\perp T_i(t) \mid W_i(t)$, where $R_i(t)$ tracks at-risk status for the primary alternating recurrent event, $T_i(t)$ is the residual primary alternating recurrent event free survival time, and $W_i(t)$ is a set of covariates that contains $Z_i(t),$ the regressors of interest.
\end{assumption}

\begin{assumption}[Sufficient regressors]\label{TimodCorrect} The restricted mean time-to-primary-event in the at-risk-population only depends on the covariates $Z_i(t)$:
$$\EE\left[\min\{T_i^{(1)}(t),\tau\}|Z_i(t),W_i(t), R_i(t)=1\right]=\EE\left[\min\{T_i^{(1)}(t),\tau\}|Z_i(t), R_i(t)=1\right].$$
\end{assumption}

A consequence of this assumption is that our POs can be fully modeled by covariates $Z_i(t)$, with no reliance on $W_i(t)$. 


\begin{theorem}[IPW corrects the selection into the at-risk set]
Under Assumptions \ref{conditionalIndependence}, \ref{TimodCorrect}, and $0<\pi_i(t)<1$, the following holds:
\begin{align}
\EE\left\{\frac{R_i(t)}{\pi_i(t)} PO_i^\tau(t)\bigg| Z_i(t)\right\} = \EE\big[\min (T_i(t), \tau )|Z_i(t)\big].
\end{align}
\end{theorem}\label{prop1}
See Section \ref{proofProp1} in the Supplementary Materials for a detailed proof.

Hence, for each individual, if we can consistently estimate $\dfrac{R_i(t) {PO}^\tau_i(t)}{P\{R_i(t) = 1|W_i(t)\}}$, this term will have the same limiting expectation as the random variable $\min \{T_i(t), \tau\}$ given $Z_i(t)$, $i = 1, 2, \ldots n(t)$. 
In Section \ref{s:RF}, we describe a random forest estimate, $\hat{\pi}(t)$ for $P\{R_i(t) = 1|W_i(t)\}$ and conditions where these estimates are consistent. Under a linear regression model specification of the target $\EE[\min \{T_i(t), \tau\} \mid Z_i(t)]=\beta^\top Z_i(t)$, the estimates $\hat \beta$ may then be taken as the solution to the following PO-based generalized estimating equation with homogeneous independent working variance-covariance matrix: 
\begin{equation}\label{eq:weightedGEE}
    \sum_{t = t_0}^{t_b}\sum_{ i =1}^{n(t)} \frac{R_i(t)}{\hat{\pi}_i(t)}Z_{i}(t)\{PO_i^\tau(t) - \beta^\top Z_{i}(t)\} = 0.
\end{equation}
Depending on the form of the censored longitudinal data structure selected, an analyst may select other variance-covariance structures such as the Toeplitz or unstructured variance-covariance matrix \citep[e.g.,][]{Xia:StatMed:2019:regression}. In this paper we use an unstructured variance covariance estimates along with standard errors found via the sandwich estimator.
Hereafter, we refer to this model as the pseudo-observation alternating recurrent event IPW and random forest GEE model, or PAIR-GEE for short; other non-identity link functions can be used as appropriate depending on the intended target regression form.


%% file: 4RF.tex
We observe pairs $\{R_i(t), W_i(t)\}$, where $R_i(t)$ is at-risk indicator for the primary event type and $W_i(t)$ is a set of $p$-many potentially time-varying and correlated covariates measured at time $t$, as well as $t$ itself. Here we describe how to obtain estimates, $\hat{\pi}_i(t)$, for $P\{R_i(t) = 1|W_i(t)\}$ using a random forest algorithm. These algorithms are pervasive in machine learning literature, although very few algorithms have been described in a way that is applicable for the longitudinal $R_i(t)$, $t\in \mathcal{T}_i$. Our general approach follows that of \citet{hrfRef} with algorithmic details specific to constructing $\hat{\pi}_i(t)$ given below. As such, we benefit from similar advantages of being less parametric than competitor GEE regression models for dependent binary responses, while capturing complex, nonlinear associations between predictors and outcomes. In the below, we lean on random forest jargon, which we explain in the context of our problem as we go through the algorithm.

The predictors $W_i(t)$ are allowed to be highly collinear, and should contain $Z_i(t)$ as well as a large number of event history variables observed prior to $t$ that are some of the most helpful predictors when event times are correlated \citep{Chapter1}. Some examples of event history variables known at $t$ include time since last event, average times-to-event for primary and secondary event types in earlier follow-up windows, and numbers of these event types prior to $t$, as well as $t$ itself. If event history data is not available at $t=0$, the random forest algorithm is able to proceed, broadening the predictor pool as $t$ moves to later follow-up windows. We will also explore the use of reflective history predictors similar to those described in missing data literature for longitudinal analyses with non-monotone dropout \citep{TchetgenTchetgen}, where these predictors use either future or past information relevant to $t$, but not both, depending on which temporal direction has more longitudinal measurements available. Such longitudinal missingness patterns are fully observed when analyzing retrospective data; thus covariates that may perfectly predict $R_i(t)$ after $t$, are available for analysis. For further discussion, see Supplemental Materials Section \ref{supp:histCovs}. Once these predictors are constructed, the random forest algorithm starts by repeating Steps 1 and 2 $B$-many times:
\begin{itemize}
    \item[Step 1)] Two-stage bootstrap sample: 
        Sample with replacement from the set of unique subjects, $i = 1, 2, \ldots n$. For each sampled subject (indexed by $\ell$) in the bootstrap dataset, randomly select one follow-up window $t_{j_\ell}\in \mathcal T_\ell$. Label the corresponding selected data pairs,  $\{R_\ell(t_{j_\ell}), W_\ell(t_{j_\ell})\}$.
        Subjects in the bootstrap sample are called \say{bagged, } with the remaining subjects referred to as \say{out-of-bag.}
    \item[Step 2)] Grow a tree:\begin{enumerate}
        \item[a.)] From the $p$ possible covariates, sample $m< p$ covariates. $m$ is oftentimes a user-selected value, but if no value is specified, many software programs default to $\sqrt{p}$ or $\frac{p}{3}$. For each $s\in \{1, \ldots m\}$, consider a grid of all possible dichotomizations (or indicator functions) of selected covariates, denoted $\{\bD_1, \bD_2, ... \bD_m\}$, where $\bD_s$ is a grid of all observed values of selected covariate $D_s.$
        \item[b.)] Consider the values of the objective function created by splitting the data into two groups based on each potential dichotomization. In our case, the outcome of interest is a Bernoulli random variable; we consider the objective function known as the Gini cost function (for further details see Chapter 4 of \citet{breiman2001random}). Bin the data into two ``daughter" nodes that reflect the split minimizing the objective function.
        \item[c.)] Recursively make splits in the daughter nodes until one of three hyperparameters is met: (1) a minimum number of observations is found in a daughter node, (2) the maximum number of splits as specified by the analyst has been made, or (3) no further reduction in the objective function can be made with more splits.
    \end{enumerate}
\end{itemize}
Predictions from tree $b$ for individual $i'$ in the associated out-of-bag sample are made by following the split decisions that correspond to covariate pattern $W_{i'}$, and averaging the at-risk indicators found in the terminal node that $i'$ belongs to. When all $B$ trees are grown, random forest predictions for individual $i'$ averages the predictions arising from trees where $i'$ is out-of-bag, allowing fitted values to remain independent from the tree-growing stage. 

There are two main advantages to this forest. First, due to the two-stage bootstrapping scheme, subjects with many measurements are as likely to be included in the sampling step as subjects with few measurements. This feature contributes to model stability in the presence of censoring, as individuals who are censored early have their observed data frequently included. Second, out-of-bag data remain independent of the bagged data, yielding simple fitted values where proofs involving consistency of independently distributed outcomes from \citet{breiman2001random} may still be applied.






%% file: 6simStudy.tex
 The goals of our simulation study are two fold: (1) to compare the performance of PAIR-GEE models to a complete case (unweighted) analysis described in Section \ref{s:POs} and (2) to demonstrate the importance of including history covariates in estimating weights via random forest. For goal (1), we report coverage probabilities of 95\% confidence intervals, average standard error, empirical standard error, bias and percent bias of estimates. 
 For goal (2), we define seven useful history variables, discussed in Supplemental Materials Section \ref{supp:histCovs}. 

We generate correlated piece-wise exponential alternating recurrent event times ($\rho\approx0.80$)  using a Gaussian copula approach similar to that described by \citet{Xia:StatMed:2019:regression} for a binary predictor, $Z = 0, 1$. Independent censoring was generated by $\min\{12, L_i\}$, where $L_i \sim \text{Laplace}(\mu = 12, b = 1.5)$, which corresponds to roughly 30\% of observations censored before the final time point. The censored longitudinal data structure takes the form $t \in \{0, 1, 2, \ldots 11\}$ months with $\tau = 1$ month. Piece-wise hazards at times $[0, 6), [6,12]$ are $\{1/2 , 1/4 \}$ for $Z= 0$ and $ \{1/3, 1/6\}$ for $Z = 1$. 
Our data follow:
\begin{align*}
E[\min\{T_i(t), \tau = 1\}] = & 0.714 \times I(Z_i=1)I(t<6) +  0.631\times I(Z_i=0)I(t< 6) + \\
& + 0.826 \times I(Z_i=1)I(t\geq 6) + 0.764 \times I(Z_i=0) I(t \geq 6).
\end{align*}
so that the 1-month restricted means for $t < 6$ are  $\{ 0.631, 0.714\}$ for $ Z = 0$ and  $Z=1$, respectively, and  for $t\geq 6$, the 1-month restricted means are $\{ 0.764, 0.826\}$ for $ Z = 0$ and  $Z=1$, respectively.
Figure \ref{supp:fig:sampSubj} in Supplemental Materials displays six sample subjects that follow the various hazards of alternating recurrent events in the correlated case. Subjects with $Z_i = 0$ have both more primary events and less availability at window start times $t \in \mathcal{T}$ in the complete case analysis. Figure \ref{supp:fig:simReplicate} in Supplemental Materials shows a representative simulated dataset and the empirical restricted means over time using an unweighted GEE approach that ignores the issue of selection into the risk set. An additional simulation, considering considering independent gap-times and continuous covariates, is included in Supplementary Materials Section \ref{supp:simStudy:ind} with results displayed in Table \ref{supp:tab:censRes} and Figure \ref{supp:beta_violin_censored_independent}.

\begin{figure}
\centerline{%
\includegraphics[width=.85\textwidth]{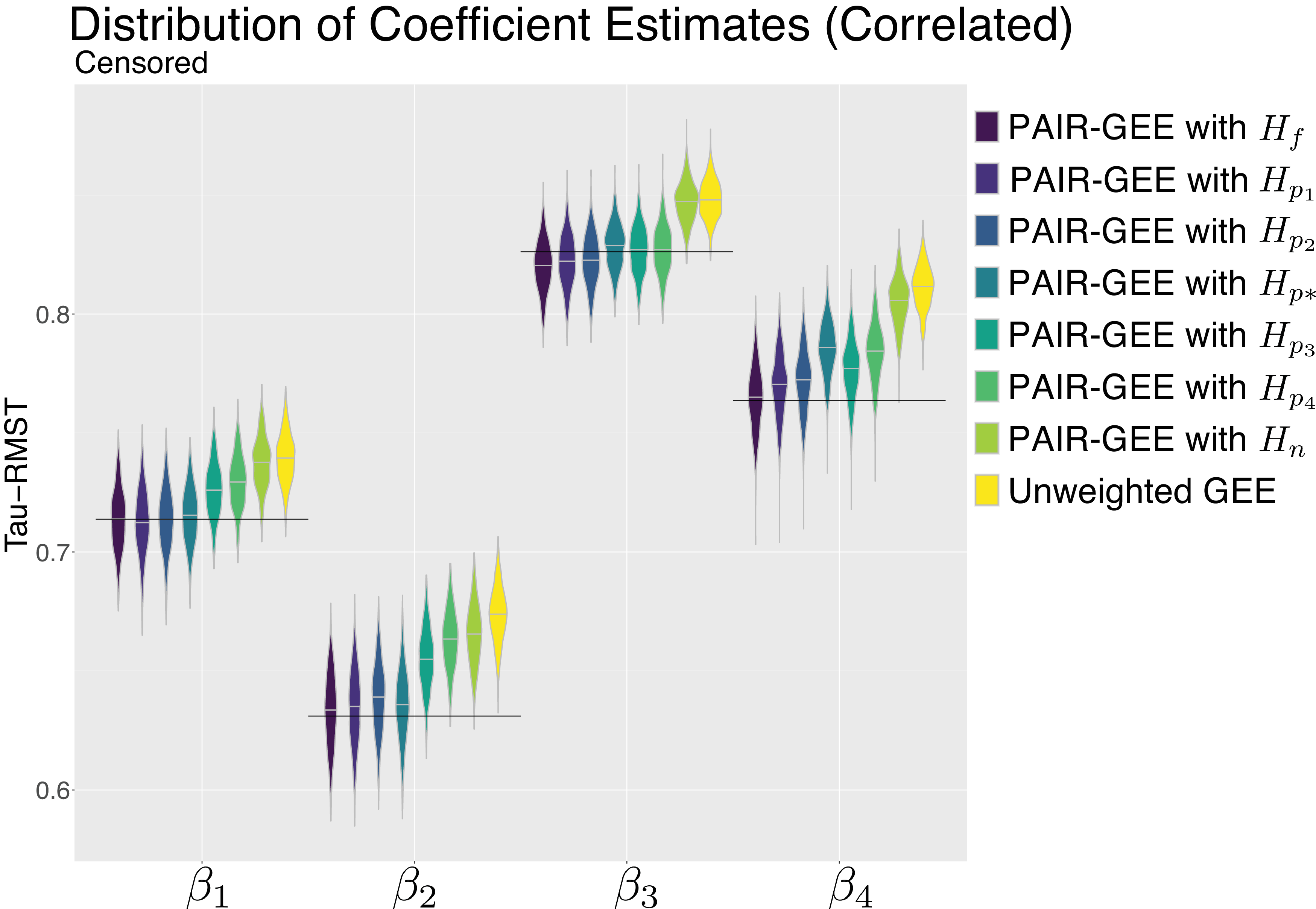}}
\caption{Violin plots of censored point estimates seen in simulation with censoring applied ($500$ replicates, $n=750$). Coefficients $\beta_1$ and $\beta_2$ reflect those with coefficients $Z_i=1$ and $Z_i = 0$ before $t = 6$ and coefficients $\beta_3$ and $\beta_4$ after and including $t = 6$, respectively. Features included in the random forest algorithm for estimating PAIR-GEE weights involve the seven history covariates described in Section \ref{s:RF} with varying degrees of availability that decreases from left to right for each displayed coefficient. Here, $\bH_f$ indicates the full year of patient history was available prior to $t=0$. $\bH_{p_1}$ and $\bH_{p_2}$ indicate that the patient history included information on the last two or one alternating recurrent events, respectively, if these occurred within the prior year. $\bH_{p*}$ indicates that no historical information prior to $t=0$ was available and that reflexive history covariates (as described in Section \ref{s:RF}) were used. $\bH_{p_3}$ indicates that complete history was available on 50\% of individuals, but was otherwise unavailable. $\bH_{p_4}$ indicates no availability of history covariates prior to $t=0$. $\bH_n$ indicates no history covariates were included in the random forest algorithm.
}\label{fig:SimRes}
\end{figure}

The seven history covariates described in Section \ref{s:RF} were included in the random forest algorithm for estimating PAIR-GEE weights. The following notation is used to indicate the degree of  of historical information available in the results shown in Figure \ref{fig:SimRes}, Table \ref{tab:CorCensoredRes} and Supplemental Table \ref{tab:CorUncensoredRes}.
$\bH_f$ indicates a full year of patient history was available prior to $t=0$. $\bH_{p_1}$ and $\bH_{p_2}$ indicate that the patient history included information on the last two or one alternating recurrent events, respectively, if these occurred within the prior year. $\bH_{p*}$ indicates that no historical information prior to $t=0$ was available and that reflexive history covariates (as described in Section \ref{s:RF}) were used. $\bH_{p_3}$ indicates that complete history was available on 50\% of individuals, but was otherwise unavailable. $\bH_{p_4}$ indicates no availability of history covariates prior to $t=0$. $\bH_n$ indicates no history covariates were included in the random forest algorithm.

The empirical distribution of coefficient estimates across simulation replicates for the censored setting is displayed in Figure \ref{fig:SimRes}, with numeric results displayed in Table \ref{tab:CorCensoredRes}. Uncensored results are displayed in Supplementary Materials Table \ref{tab:CorUncensoredRes} and Figure \ref{supp:fig:uncensored_beta_plot}. As expected, the unweighted GEE analysis gives the poorest performance in terms of bias and coverage. Any use of covariate information in estimating IPW weights reduces bias. However, success varies greatly depending on the amount of historical information. Thus, results emphasize the importance of seeking out as much historical covariate information as possible in estimating these weights.

Whereas methods with high degree of historical covariate information ($\bH_f$, $\bH_{p_1}$, $\bH_{p_2}$ and $\bH_{p*}$) gave excellent results in terms of bias and 95\% confidence interval coverage, as historical information becomes less available ($\bH_{p_3}$, $\bH_{p_4}$, $\bH_{n}$) performance deteriorates.
One exception was in estimating $\beta_4$, which was influenced by an abrupt change in hazard after $t\geq 6$. All methods struggled in terms of bias and 95\% coverage probability, though the magnitude of under coverage varies drastically (coverage probability of $0.928$ for PAIR-GEE with $\bH_f$ weights to $0.008$ for the unweighted GEE method). This phenomenon is further evaluated via Figure \ref{supp:fig:win_by_win_bias}, which shows the window-by-window bias for all PAIR-GEE models and the unweighted model; for $t$ near $6$, PAIR-GEE models that rely on history covariates have varying, non-zero levels of bias, while somewhat stabilizing near zero by the final window start time, $t=11$. For the PAIR-GEE model with $\bH_n$ and the unweighted model, this does not happen, but rather after $t=0$, bias consistently remains around $0.05$. 

A highlight of the simulation results is the strong performance of the reflective covariate $\bH_{p*}$ that uses either past or present information only on study time relative to time $t$, but not both, depending on which direction had more measurements available for construction of history covariates. This predictor greatly outperformed the historical covariates of type $\bH_{p_4}$, that had the same amount of historical information available, but only used information prior to $t$ in estimating the IPW weights. 

\begin{table}[!htbp] \centering 
  \caption{ A table displaying results from simulation in the case where primary recurrent events have correlation of $\rho = 0.8$, and secondary recurrent events are independent of each other. The hazard of primary alternating recurrent event completely determines the hazard of the alternating secondary recurrent event, approximately 30\% of subjects are censored before the final follow-up window, and results come from 500 replicates of $n=750$ each. } 
  \label{tab:CorCensoredRes}
\begin{tabular}{@{\extracolsep{1pt}} lcrccccc} 
Model & Covariate &  $\hat\beta-\beta$ & $|\hat\beta-\beta| / \beta$ & Coverage & $\hat{SE}$ & ESD & $\hat{SE}$/ESD\\\hline 
\hline \\[-1.8ex]
PAIR-GEE, with $\bH_f$ & $\hat\beta_1$ &  0.000 & 0.000 & 0.934 & 0.013 & 0.014 & 0.963 \\ 
 & $\hat\beta_2$ & -0.002 & 0.003 & 0.934 & 0.015 & 0.016 & 0.919 \\ 
 & $\hat\beta_3$ &  0.006 & 0.007 & 0.946 & 0.011 & 0.011 & 1.046 \\ 
 & $\hat\beta_4$ & -0.001 & 0.001 & 0.928 & 0.013 & 0.014 & 0.929 \\ 
 \hline 
PAIR-GEE, with $\bH_{p_1}$ & $\hat\beta_1$ &  0.001 & 0.002 & 0.932 & 0.014 & 0.015 & 0.925 \\ 
 & $\hat\beta_2$ & -0.003 & 0.005 & 0.918 & 0.015 & 0.016 & 0.895 \\ 
 & $\hat\beta_3$ &  0.004 & 0.005 & 0.950 & 0.011 & 0.011 & 1.018 \\ 
 & $\hat\beta_4$ & -0.006 & 0.008 & 0.884 & 0.013 & 0.014 & 0.921 \\ 
 \hline 
PAIR-GEE, with $\bH_{p_2}$ & $\hat\beta_1$ &  0.000 & 0.000 & 0.924 & 0.013 & 0.014 & 0.928 \\ 
 & $\hat\beta_2$ & -0.007 & 0.012 & 0.890 & 0.014 & 0.015 & 0.933 \\ 
 & $\hat\beta_3$ &  0.004 & 0.004 & 0.954 & 0.011 & 0.011 & 1.018 \\ 
 & $\hat\beta_4$ & -0.008 & 0.011 & 0.856 & 0.013 & 0.014 & 0.933 \\ 
 \hline 
PAIR-GEE, with $\bH_{p*}$ & $\hat\beta_1$ & -0.002 & 0.002 & 0.952 & 0.013 & 0.013 & 1.051 \\ 
 & $\hat\beta_2$ & -0.004 & 0.006 & 0.934 & 0.015 & 0.015 & 1.009 \\ 
 & $\hat\beta_3$ & -0.002 & 0.003 & 0.958 & 0.010 & 0.010 & 1.081 \\ 
 & $\hat\beta_4$ & -0.022 & 0.029 & 0.512 & 0.011 & 0.012 & 0.986 \\ 
 \hline
PAIR-GEE, with $\bH_{p_3}$ & $\hat\beta_1$ & -0.013 & 0.018 & 0.806 & 0.012 & 0.012 & 0.983 \\ 
 & $\hat\beta_2$ & -0.024 & 0.038 & 0.556 & 0.013 & 0.013 & 0.985 \\ 
 & $\hat\beta_3$ & -0.001 & 0.001 & 0.948 & 0.011 & 0.010 & 1.019 \\ 
 & $\hat\beta_4$ & -0.013 & 0.017 & 0.790 & 0.012 & 0.013 & 0.952 \\ 
 \hline
PAIR-GEE, with $\bH_{p_4}$ & $\hat\beta_1$ & -0.016 & 0.022 & 0.712 & 0.012 & 0.012 & 0.983 \\ 
 & $\hat\beta_2$ & -0.032 & 0.051 & 0.292 & 0.013 & 0.013 & 0.992 \\ 
 & $\hat\beta_3$ & -0.001 & 0.001 & 0.944 & 0.011 & 0.010 & 1.019 \\ 
 & $\hat\beta_4$ & -0.020 & 0.026 & 0.574 & 0.011 & 0.012 & 0.950 \\ 
 \hline
PAIR-GEE, with $\bH_n$& $\hat\beta_1$ & -0.024 & 0.034 & 0.414 & 0.011 & 0.011 & 0.973 \\ 
 & $\hat\beta_2$ & -0.034 & 0.054 & 0.224 & 0.012 & 0.013 & 0.953 \\ 
 & $\hat\beta_3$ & -0.021 & 0.025 & 0.328 & 0.009 & 0.009 & 1.024 \\ 
 & $\hat\beta_4$ & -0.041 & 0.054 & 0.026 & 0.010 & 0.010 & 0.962 \\ 
 \hline 
Unweighted GEE & $\hat\beta_1$ & -0.026 & 0.036 & 0.328 & 0.011 & 0.011 & 1.019 \\ 
 & $\hat\beta_2$ & -0.043 & 0.067 & 0.058 & 0.012 & 0.012 & 1.000 \\ 
 & $\hat\beta_3$ & -0.022 & 0.027 & 0.296 & 0.009 & 0.008 & 1.089 \\ 
 & $\hat\beta_4$ & -0.047 & 0.062 & 0.008 & 0.010 & 0.009 & 1.011 \\ 

\hline \\[-1.8ex] 
\end{tabular} 
\end{table}


%% file: 7DataApplication.tex
\label{s:data-app:TBI}

The 6-month CareQOL study \citep{CareQoLApp} randomized $n=257$ caregivers of traumatic brain injury patients at baseline to receive, or not receive self-care push notifications from the CareQOL app, which tailored the message content by previous levels of physical activity, and sleep minutes measured by FitBit devices along with previous levels of anxiety, strain, and depression based on ecological momentary assessments (health-related quality of life, HRQOL). Other demographic information (ethnicity, age, race, sex) and the caregiver's patient level of independence were recorded by study personnel. All study participants had weekly self-reported HRQOL measures of depression ($t$-score). Subjects with a weekly $t$-score two points above baseline are labeled as in a depressive state. Hence our alternating recurrent events are entering (primary) and exiting (secondary) a depressed state, respectively. The findings from this study suggest that self-care push notifications did not enhance outcomes (somewhat worse main effect), with benefits observed among participants with active engagement with the app \citep{CareQoLApp}. Here we take a deeper dive and analyze the weekly data over the 6-month period. Using our proposed method, we are able to analyze transitions in and out of the depressed state to interpret what may have contributed to the non-improvement of the end-of-study HRQOL outcomes.

We apply our proposed method to 12 weeks of follow-up of the CareQOL cohort with $\tau = 2$ weeks and $t\in \mathcal{T} = \{0, 2, 4, 6, 8, 10\}$ weeks. We fit the following marginal model using GEE, with and without random forest IPW weights, supplying the random forest algorithm with $\bH_p*$ variables as described in Section \ref{supp:histCovs} of the Supplementary Materials as well as other patient characteristics:
\begin{align*}
E[\min\{T_i(t), \tau\}] &= \beta_0 + \beta_1 I(\text{Ethnicity}_i = \text{Non-Hispanic}) + \beta_2 \text{Age}_i + \beta_3 I(\text{Race}_i= \text{White})\\
& \quad +\beta_4I( \text{Male}_i) + \beta_5I(\text{Caregiver's TBI Patient}_i = \text{Independent})\\
& \quad +\beta_6I(\text{Caregivers' TBI Patient}_i = \text{Partially Independent}) \\
& \quad + \beta_7 \text{Sleep Minutes}_i(t)+ \beta_8 \text{Steps}_i(t) + \beta_9 I(\text{Push Notifications}_i).
\end{align*}

\begin{figure}
\centerline{%
\includegraphics[width=\textwidth]{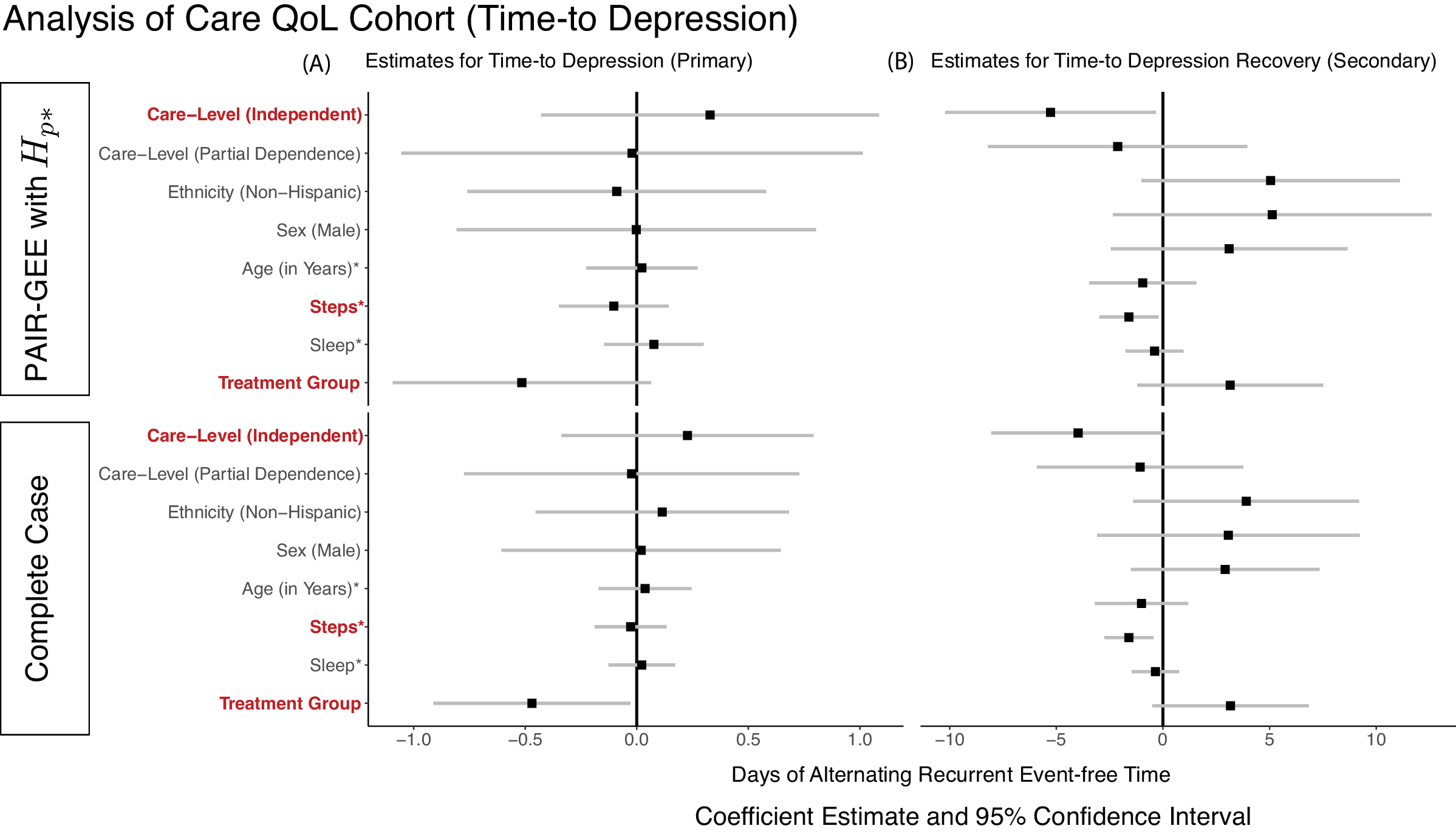}}
\caption{Point estimates and corresponding 95\% confidence intervals from the CareQOL data analysis. The left and right panels correspond to the estimates of $2$-week restricted mean time-to-entering and exiting depressed states, respectively. Significant coefficients, or coefficients that change significance between methods are displayed in a bold, red font. Covariates that were standardized prior to inclusion in the model are denoted with an asterisk. Our analysis finds that after adjustment for common covariates, being in the intervention group was no longer significantly associated with a decrease in the number of days till entering a depressive state.}\label{fig:careQoL}
\end{figure}

Point estimates of coefficients are displayed in Figure \ref{fig:careQoL}. When using the PAIR-GEE method that correctly accounts for selection into the risk set, push notifications were associated with $0.51$ fewer days of depression-free status for every two weeks of follow-up (95\% CI $-1.095$ to $0.065$, p-value $0.082$), adjusting for other covariates. This corroborates with the main study findings where the end-of-study depression measure was not enhanced. Notably, when patients who are more susceptible to be in a depressive state are insufficiently accounted for, as in the unweighted analysis, the treatment effect rises to statistical significance ($0.47$ fewer days of depression-free status per two weeks of follow-up, 95\%  CI $-0.028$ to $-0.912$, p-value $0.037$).

For both weighted and unweighted methods, time-to depression recovery (secondary event type) shortens as standardized step count increases (Figure \ref{fig:careQoL}, Panel (B), approximately $1.59$ fewer days depressed per two week period for every $\hat{\sigma} = 3,763$ steps) while adjusting for other covariates. Additionally, the PAIR-GEE model detects a much stronger association between caregivers of independent TBI patients and time-to depression recovery (approximately $5.27$ days fewer than caregivers for fully dependent TBI patients) than the complete case analysis (roughly $3.98$ fewer days).


%% file: 8Discussion.tex
We presented a novel IPW pseudo-observation based approach to estimating $\tau$-restricted mean time-to-event for the primary event type using alternating recurrent event data, and discussed different covariates to be used in the estimation of these weights. In settings with high quality patient history information and correlated alternating recurrent events, our model is able to estimate $\tau$-restricted means while correcting for the bias caused by implicit selection of subjects into the at-risk population that a complete case analysis suffers.

Of note is the weakness of the proposed model that does not use any history coefficients ($\bH_n$); it performs almost as poorly as the unweighted model. We make three general recommendations on the use of history variables: (1) whenever possible, full history covariates should be obtained,  unless when the hazard of the primary event type is constant over time; (2) if the full history is not available, at least the timings of the last alternating recurrent event pair within a subject is sufficient; and (3) in the absence of any historical information, reflective history that accounts for missingness pattern at $t> t_0$  may be substituted. One of the benefits of our method is that the analysis of alternating recurrent events is agnostic to different primary event definitions. For example, the definition of $T_i(t)$ may be time-to depressive state after time $t$, but we also may define it as time-to exiting the depressive state after time $t$, that is, a reversal of primary and secondary event definitions.

In simulations, we noted the consistent under coverage of the $\tau$-restricted mean after a change in hazard was observed. This leads to interesting questions about how one might adjust for changes in the health state when calculating patient-specific history covariates. Important work in the future may consider kernel density-type calculations of history covariates to adjust for more recent patient specific histories, rather than treating all historical events as equally important at time $t$. As health data become richer and more temporally granular due to electronic health records and fitness trackers, these methodological refinements will be essential.